\documentclass[runningheads]{llncs}
\usepackage{graphicx}
\graphicspath{{./img/}}
\usepackage{array}
\newcolumntype{P}[1]{>{\centering\arraybackslash}p{#1}}
\newcolumntype{M}[1]{>{\centering\arraybackslash}m{#1}}

\usepackage[ruled,linesnumbered]{algorithm2e}
\usepackage{color}
\usepackage{colortbl}

\usepackage{amssymb}

\let\oldnl\nl
\newcommand{\nonl}{\renewcommand{\nl}{\let\nl\oldnl}}
\DeclareMathAlphabet      {\mathbfit}{OML}{cmm}{b}{it}

\SetCommentSty{mycommfont}

\usepackage{enumitem}
\setlist{nosep}

\begin{document}
\title{Orchestrated Co-Scheduling, Resource Partitioning, and Power Capping on CPU-GPU Heterogeneous Systems via Machine Learning}

\titlerunning{Orchestrated Co-Scheduling, Resource Partitioning, and Power Capping}

\author{
Issa Saba \and
Eishi Arima \and
Dai Liu \and
Martin Schulz}

%
\institute{Technical University of Munich, Garching, Germany\\
\email{\{issa.saba,eishi.arima,dai.liu,martin.w.j.schulz\}@tum.de}
\vspace{-5pt}
}

%
%
%
%
\maketitle              
\begin{abstract}
CPU-GPU heterogeneous architectures are now commonly used in a wide variety of computing systems from mobile devices to supercomputers. 
Maximizing the throughput for multi-programmed workloads on such systems is indispensable as one single program typically cannot fully exploit all avaiable resources. 
At the same time, power consumption is a key issue and often requires optimizing power allocations to the CPU and GPU while enforcing a total power constraint, in particular when the power/thermal requirements are strict. 
The result is a system-wide optimization problem with several knobs. In particular we focus on
(1) co-scheduling decisions, i.e., selecting programs to co-locate in a space sharing manner; (2) resource partitioning on both CPUs and GPUs; and (3) power capping on both CPUs and GPUs. 
We solve this problem using predictive performance modeling using machine learning 
in order to coordinately optimize the above knob setups. 
Our experiential results using a real system show that our approach 
achieves up to 67\% of speedup 
compared to a time-sharing-based scheduling with a naive power capping that evenly distributes power budgets across components.

\end{abstract}
\vspace{-22pt}

\section{Introduction}\label{intro}
\vspace{-3pt}

Heterogeneous CPU-GPU architecture are now broadly used in a wide variety of computing systems, including mobile devices, PCs, datacenter servers, and HPC systems. 
For instance, over 160 out of the 500 top-class supercomputers are now GPU-accelerated systems (as of Jun 2022)~\cite{top500}. 
This trend is driven by the end of Dennard scaling~\cite{dennard}, i.e., the exponential growth of single-thread performance in microprocessors had ceased, and the industry rather shifted toward thread-level parallelism and heterogeneous computing using domain specific accelerators~\cite{dennard2}. 
GPUs are one of the most commonly used accelerators due to their wide range of application areas, including image processing, scientific computing, artificial intelligence and data mining.

As computing systems are becoming more powerful and more heterogeneous using a wide variety of resources, it also becoming more difficult to fully utilize the entirety of compute resources by one single application. 
One reason behind the trend is that it is not always easy to identify a large enough fraction of a code that can be ported to GPUs (or any other accelerator) while balancing loads across all the processing units (CPU, GPU, or any accelerators). 
Further, the scalability of applications inside of a chip can be limited by various factors such as intensive memory accesses and shared resource contentions, which can induce a significant waste of compute resources.

Therefore, \textbf{co-scheduling}, i.e., co-locating multiple processes in a space sharing manner, is a key feature to mitigate resource wastes and to maximize throughput on such systems, if the processes are complimentary in their resource usage.
To achieve the latter, a sophisticated mechanism to \textbf{partition resources} at each component and allocate them accordingly to co-scheduled processes is indispensable. 
Recent commercial CPUs and GPUs support such resource partitioning features: (1) one can designate physical core allocations to co-scheduled processes on CPUs; and (2) GPUs have begun to support hardware-level resource partitioning features for co-locating multiple processes --- one example is NVIDIA's Multi-Instance GPU (or MIG) feature that is supported in the \textit{most recent} high-end GPUs to enable co-locating multiple programs at the same chip while partitioning it at the granularity of GPC~\cite{mig2}. 

Meanwhile, as power (or energy) consumption is now a first-order concern in almost all the computing systems from embedded devices to HPC systems~\cite{hpc-power3,hpc-power,hpc-power2}, performance optimizations for modern computing systems, including co-scheduling, must consider power optimization and in most cases also hard power limits or constraints. 
Once a power constraint is set on a system, the power budgets must be distributed to components accordingly so as to maximize the performance while keeping the constraint. 
To realize such a mechanism, modern CPUs and GPUs now support \textbf{power capping} features that set a power limit at the granularity of chip (or even at a finer granularity for some hardware).

Driven by the above trends, this paper explicitly targets the combination of co-scheduling, resource partitioning, and power capping on CPU-GPU heterogeneous systems, and provides a systematic solution to co-optimize them using a machine-learning-based performance model as well as a graph-based scheduling algorithm. 
Our model takes both application profiles and hardware knob states into account as its inputs and returns the estimated performance of the co-located applications as the output. 
More specifically, the profiles are based on hardware performance counters, and the hardware knob states include resource partitioning and power capping on both the CPU and GPU. 
We use this performance model to estimate the best performance of different hardware setups for a given application pair, which is used to determine the best co-scheduling pairs in a graph-based algorithm, i.e., Edmonds' algorithm~\cite{Edmond}.

The followings are the major contributions of this paper:
\begin{enumerate}
	\item We comprehensively and systematically optimize (1) co-scheduling pair selections, (2) resource partitioning at both CPU and GPU, and (3) power budgeting on both CPU and GPU, using a real CPU-GPU heterogeneous hardware platform. 
	\item We define an optimization problem and provide a systematic solution to select the best job pair and the best hardware setups including resource partitioning and power capping on CPU/GPU.
	\item We develop a machine-learning-based performance model that takes both the characteristics of the co-located applications and the hardware states 
	(including partitioning and power capping on CPU/GPU) into account. 
	\item We solve the optimization problem by using the above performance model building on the graph-based Edmonds' algorithm. 
    \item We quantify the benefits of our approach by using a real hardware, and show that we improve the system throughput by 67\% compared to a time-sharing-based scheduling with a naive power capping that evenly distributes the power budgets across the CPU and GPU.
\end{enumerate}


\vspace{-5pt}
\section{Related Work}
\vspace{-5pt}

Ever since multi-core processors appeared on the market, co-scheduling, resource partitioning, and power capping have been studied. However, ours is the first work that covers all of the following aspects simultaneously: (1) targeting CPU-GPU heterogeneous systems; (2) comprehensively co-optimizing co-scheduling pair selections, resource partitioning, and power capping, using machine-learning-based performance modeling and a graph-based algorithm; and (3) quantifying the benefits using a real hardware that is capable of both resource partitioning and power capping at both the CPU and the discrete GPU.

M. Bhadauria et al. explored co-scheduling multi-threaded programs in a space sharing manner using a multi-core processor~\cite{cosh-ics10}. 
S. Zhuravlev et al. focused on the shared resource contention across co-located programs on multi-core processors and proposed a scheduler-based solution to mitigate the interference~\cite{cosh-asplos10}. 
R. Cochran et al. proposed Pack \& Cap that optimizes the scale of multi-threaded applications via the thread packing technique while applying power capping~\cite{cpu-cosh-powcap}. 
Then, H. Sasaki et al. provided a sophisticated power-performance optimization method that coordinates the thread packing technique and DVFS for multi-programmed and multi-threaded environments~\cite{cpu-cosh-powcap2}. 
These seminal studies provided insightful ideas, \textit{however they did not target CPU-GPU heterogeneous systems.}

Few studies looked at the combination of co-scheduling
and power capping on \textit{CPU-GPU heterogeneous systems}. Q. Zhu et al. worked on the combination of job scheduling and power capping for integrated CPU-GPU systems~\cite{cpu-gpu-cosh}, but they did not cover the following aspects: resource partitioning inside of CPU/GPU; and co-scheduling multiple processes on the GPU in a space sharing manner. 
R. Azimi et al. developed a framework called PowerCoord that allocates power budgets to components on CPU-GPU heterogeneous systems for co-scheduled workloads~\cite{power-coord}, but their work did not target adjusting the resource partitions as well. 
\textit{Recent hardware advances (e.g., NVIDIA MIG feature~\cite{mig-icppw,mig2}) made it possible to apply both the process co-location and resource partitioning on both CPUs and GPUs, which opened up new optimization opportunities. }

There have been several studies that utilize machine learning (including linear regression) for performance/power modeling in the literature. 
B. Lee et al. utilized the linear regression to predict performance for CPUs~\cite{linear-regression}. 
E. \"{I}pek et al. conducted microarchitectural design space explorations using a neural network~\cite{nn-design}. 
B. Barnes et al. proposed a statistical approach to predict performance of parallel applications~\cite{lr-application2}. 
H. Nagasaka et al. constructed a power consumption model for GPUs that is based on the linear regression and hardware performance counters~\cite{lr-application3}. 
Beyond these pioneering studies, machine-learning-based approaches have been utilized also for more complicated system design and optimization purposes such as: clock frequency setups at both CPU and GPU at the same time on a CPU-GPU integrated chip~\cite{lr-application4}; power capping setups on CPU, DRAM, and NVRAM~\cite{lr-application5}; coordination of thread/page mapping and prefetcher configurations~\cite{numa-prefetch}; and CPU-GPU heterogeneous chip designs in the industry~\cite{hetero-design}. 
We follow the literature and utilize a neural network that is tailored to solve our new problem.

\section{Motivation, Problem, and Solution Overview}

\subsection{Motivation: Technology Trends}
Setting a power cap on a processor is a crucial feature and is now supported on a variety of commercial CPUs and GPUs. 
One prominent use case for this feature is to protect a chip from overheating and, instead of having to be conservative, to adjust the needed settings to the machine environment such as the cooling facility. 
Another prominent use case is enabling a hierarchical and cooperative power budgeting across components or compute nodes while keeping a total power constraint, which has been widely studied from standalone computers to large-scale systems, including datacenters and supercomputers~\cite{hpc-power3,hpc-power,hpc-power2}. 
In our work, we target CPU-GPU heterogeneous computing systems (or nodes) and optimize the power cap setups on both CPU and GPU in order to maximize performance under a total node power constraint. 

As compute nodes are becoming fatter and systems more heterogeneous, it is also becoming more difficult to fully utilize an entire node's resources by one single process. 
For instance, compute resources are typically under utilized for memory-bound applications, while memory bandwidth is often wasted for compute-bound applications. Further,
some applications are suitable for running on GPUs, but others are not. 
To improve resource utilization, mixing different kinds of processes and co-scheduling (or co-locating) them on the node at the same time while setting resource allocations accordingly at both the CPU and GPU is a desired feature.

\subsection{Problem Definition}\label{prob-def}

\begin{figure}[t]
  \centering
  \includegraphics[width=0.7\linewidth]{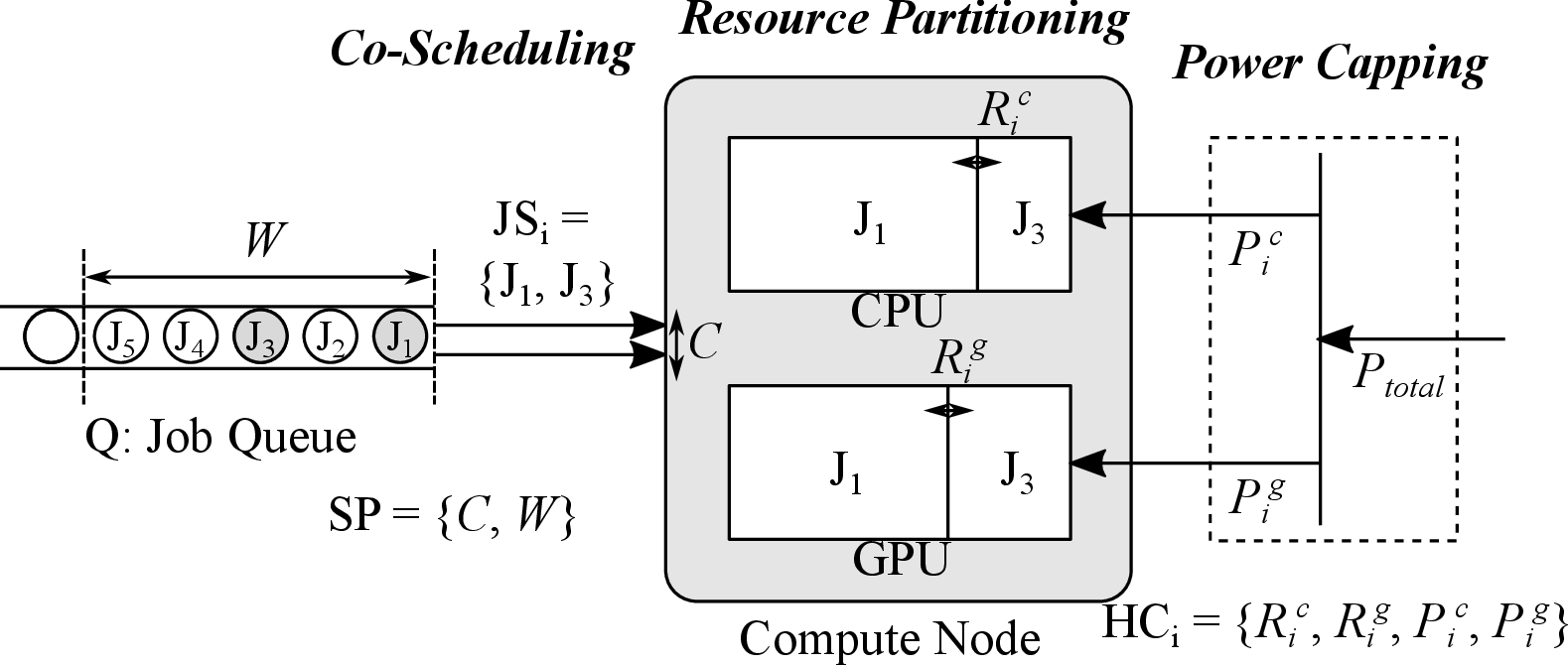}
  \caption{Problem Overview}
  \label{fig:problem}
\end{figure}

Figure~\ref{fig:problem} illustrates the overall problem we target in this paper. Here, we assume that we have one single job (or process) queue on the system ($\mathrm{Q}$). 
We convert the job queue into a list of job sets (or pairs) to co-schedule ($\mathrm{JS_1}, \mathrm{JS_2}, \cdots$). 
Note these jobs are selected from inside of the window ($W$) on the queue.  
The concurrency, i.e., the maximum number of jobs launched at a time, is limited by a given parameter ($C$), and we particularly target $C=2$ in this paper, meaning that no more than 2 CPU-GPU jobs will be co-scheduled at any given time. This value was chosen as for higher values no polynomial-time algorithms for job-set selection is known.
We represent a set of these scheduling parameters as $\mathrm{SP} = \{C, W\}$. 
When launching/co-locating the $i$th job set ($\mathrm{JS_i}$), we optimize the hardware knob configurations ($\mathrm{HC_i}$), i.e., resource partitioning on CPU/GPU ($R^c_i$/$R^g_i$) as well as the power cap setups on CPU/GPU ($P^c_i$/$P^g_i$). Note, the sum of the power caps must be less than or equal than the given total power constraint $P_{total}$.

The following is the mathematical formulation as an optimization problem: 
\begin{eqnarray}
&inputs& \mathrm{Q} = \{\mathrm{J_1}, \mathrm{J_2}, \cdots, \mathrm{J_{W}}\}, P_{total}, \mathrm{SP} \nonumber\\
&outputs& \mathrm{L_{JS}} = \{\mathrm{JS_1}, \mathrm{JS_2}, \cdots\}, \mathrm{L_{HC}} = \{\mathrm{HC_1}, \mathrm{HC_2}, \cdots\} \nonumber\\
&\min& \sum_{i=1}^{|\mathrm{L_{JS}}|}CoRunTime(\mathrm{JS_i},\mathrm{HC_i})\nonumber\\
&s.t.& CoRunTime(\mathrm{JS_i},\mathrm{HC_i}) \leq SoloRunTime(\mathrm{JS_i}, P_{total}) \nonumber\\
&\ & P_i^c+P_i^g \leq P_{total}, \hspace{5pt} 1 \leq |\mathrm{JS_i}| \leq C\nonumber\\
&\ & (\forall i : 1 \leq i \leq |\mathrm{L_{JS}}| (= |\mathrm{L_{HC}}|)) \nonumber\\
&\ & \mathrm{JS_1}\cup\cdots\cup\mathrm{JS_{|L_{JS}|}} = \mathrm{Q}, \hspace{5pt} |\mathrm{JS_1}|+\cdots+|\mathrm{JS_{|L_{JS}|}}| = W \nonumber
\end{eqnarray}
The inputs are the job queue, the total power cap setup, and the set of the scheduling parameters. 
The outputs are the lists of job sets ($\mathrm{L_{JS}}$) and the associated hardware configurations ($\mathrm{L_{HC}}$). 
The objective is to minimize the sum of the co-run execution time ($CoRunTime$), each of which is a function of the co-located jobs as well as the hardware configurations. 

We take several constraints for this optimization problem into count: 
the first is the requirement that the space-sharing co-run execution should take shorter time than the time-sharing execution with exclusive solo-runs under the power cap ($SoloRunTime$) --- otherwise we should not co-schedule them. 
The second one is the power constraint, i.e., the sum of the CPU/GPU power caps must be less than or equal to the total node power cap. 
The next constraint regulates the concurrency on the system, i.e., the number of jobs in a set to be co-scheduled ($\mathrm{JS_i}$) must be less than or equal to $C$. 
The last two constraints specify that the list of job sets ($\mathrm{L_{JS}}$) must be created from the job queue ($\mathrm{Q}$) in a mutually exclusive and collectively exhaustive manner. 
Note Table~\ref{table:refs} summarizes the parameters/functions used. 


\begin{table}[t]
{
\scriptsize
\centering
\begin{tabular}{|M{0.27\linewidth}||M{0.70\linewidth}|}
    \hline
    Parameter or Function & Remarks \\
    \hline\hline
    $\mathrm{Q}$ & A list or queue of jobs: $\mathrm{Q} = \{\mathrm{J_1}, \mathrm{J_2}, \cdots, \mathrm{J_{W}}\}$ \\
    \hline
    $\mathrm{J_i}$ & $i$th job in the job list (or queue) \\
    \hline
    $P_{total}$ & The total power cap for the target computing node \\
    \hline
    $\mathrm{SP}$ & A set of scheduling parameters: $\mathrm{SP} = \{C, W\}$ \\
    \hline
    $\mathrm{C}$ & The maximum number of concurrently executed jobs \\
    \hline
    $\mathrm{W}$ & The number of scheduling targets on the job queue \\
    \hline
    $\mathrm{L_{JS}}$ & A list of job sets to be co-scheduled: $\mathrm{L_{JS}} = \{\mathrm{JS_1}, \mathrm{JS_2}, \cdots\}$ \\
    \hline
    $\mathrm{JS_i}$ & $i$th set of jobs in $\mathrm{L_{JS}}$ to be co-scheduled \\
    \hline
    $\mathrm{L_{HC}}$ & A list of hardware configurations associated with the job sets: $\mathrm{L_{HC}} = \{\mathrm{HC_1}, \mathrm{HC_2}, \cdots\}$ \\
    \hline      
    $\mathrm{HC_i}$ & The hardware configurations for $i$th job set: $\mathrm{HC_i} = \{R^c_i, R^g_i, P^c_i, P^g_i\}$
    \\
    \hline
    $R^{*}_i (* = c/g)$ & The resource partitioning setup on CPU/GPU for $i$th job set \\
    \hline    
    $P^{*}_i (* = c/g)$ & The power cap set up on CPU/GPU for $i$th job set \\
    \hline
    $CoRunTime(\mathrm{JS_i}$, $\mathrm{HC_i})$ & The total execution time when co-locating $\mathrm{JS_i}$ with $\mathrm{HC_i}$ \\
    \hline    
    $SoloRunTime(\mathrm{JS_i}, P_{total})$ & The total time when executing $\mathrm{JS_i}$ in a time-sharing manner under the total power cap ($P_{total}$); The power caps to CPU/GPU are optimized for each job execution\\
    \hline

\end{tabular}
}
\caption{Definitions of Parameters/Functions}
\label{table:refs}
\end{table}

\begin{figure}[t]
  \centering
  \includegraphics[width=0.6\linewidth]{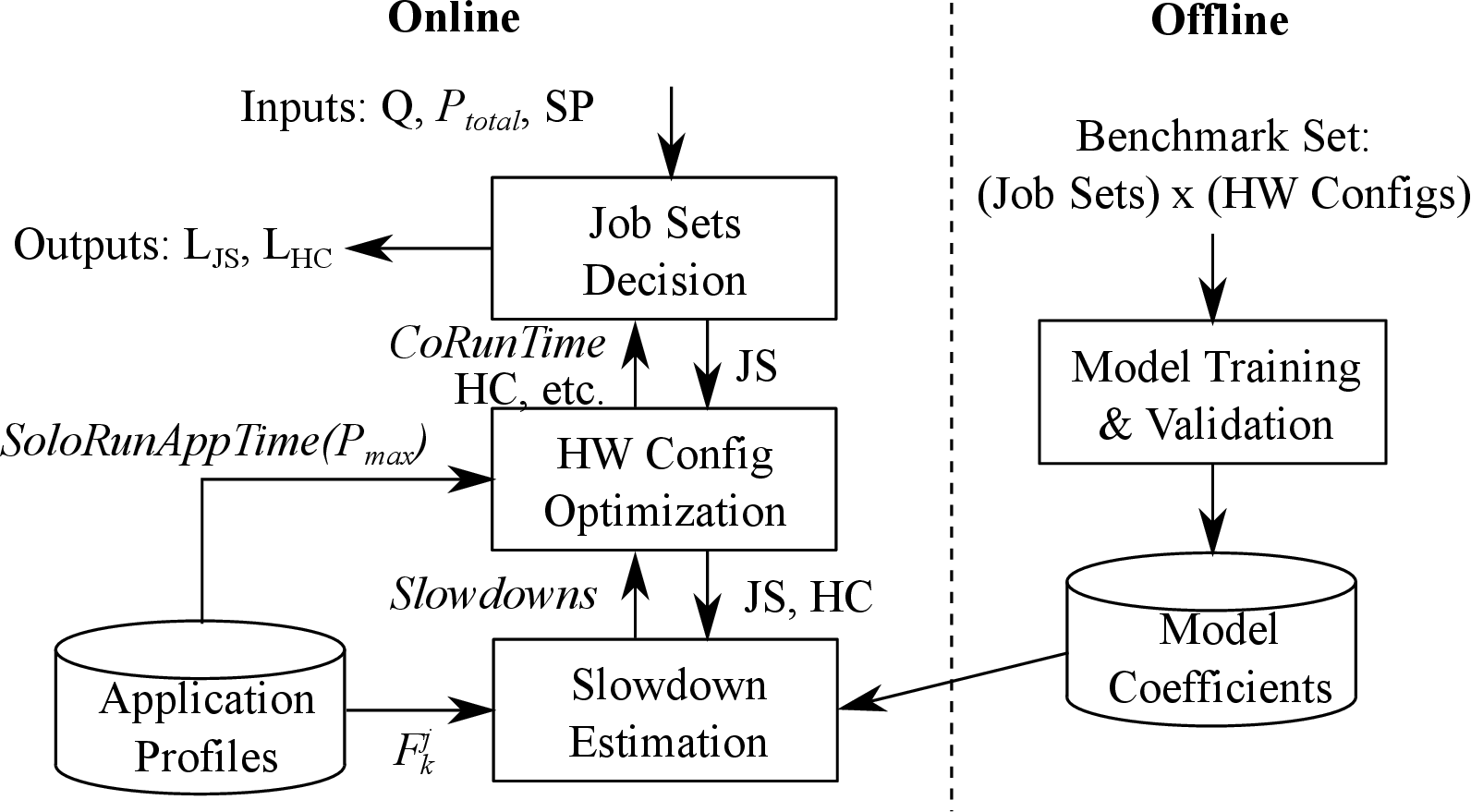}
  \caption{Workflow of Our Solution}
  \label{fig:workflow}
\end{figure}

\vspace{-8pt}
\subsection{Solution Overview}\label{solution-overview}
\vspace{-2pt}

Figure~\ref{fig:workflow} depicts the overall workflow of our approach. As shown in the figure, it consists of an offline (right) and an online part (left). 

During the offline part, we train the coefficients of our performance model, which we describe later in the paper, by using a predetermined benchmark set. 
More specifically, by executing various job sets while changing the hardware configurations, we generate a large enough number of data sets, which are used as inputs for the model training. 

During the online part, we solve the optimization problem  described in Section~\ref{prob-def}. 
This solution process consists of three parts (from top to bottom), and they work in a cooperative manner. 
We first 
determine the list of co-scheduling job sets ($\mathrm{L_{JS}}$) and return it with the associated list of optimal hardware configurations ($\mathrm{L_{HC}}$) (top part in the left figure). 
This component then communicates with the next stage (middle part in the left figure), i.e., continuously sends a temporal job set ($\mathrm{JS}$) and receives the estimated co-run execution time ($CoRunTime$) along with the optimal hardware configurations ($\mathrm{HC}$) and the solo-run time ($SoloRunTime$), which are used for the scheduling decisions. 
The component in the middle optimizes the hardware configurations ($\mathrm{HC}$) for the job set ($\mathrm{JS}$) given by the previous component. 
More specifically, it continuously sends the job set ($\mathrm{JS}$) and a temporal hardware configuration ($\mathrm{HC}$) to the third part (bottom part in teh left figure) and receives the estimated slowdowns until finding the optimal hardware configuration. 
The latter component estimates the slowdowns for the given jobs ($\mathrm{JS}$) with the given hardware configuration ($\mathrm{HC}$) by using the associated job profiles as well as the model coefficients obtained in the offline model training. 
Here, we assume that the profile of a job is collected beforehand during its first run without co-scheduling nor power capping\footnote{In case no profile is available for a job, which we do not cover in the paper, we can exclude it from the co-scheduling candidates at the first stage in the diagram and execute it exclusively without power capping while obtaining the profile for the future references.}. 
The details are described in the next section that provides also the definitions of $SoloRunAppTime(P_{max})$, $F^j_k$, and $Slowdown$ shown in the figure. 


\section{Modeling and Optimization}

\subsection{Slowdown Estimation for a Given Job Set and Hardware Setup}\label{modeling}
\subsubsection{Metric Formulations:}
We first provide simple formulations for the metrics appeared in Section~\ref{prob-def} as follows: 
\begin{eqnarray}
CoRunTime(\mathrm{JS},\mathrm{HC}) &=& \max_{\mathrm{J_j}\in \mathrm{JS}}CoRunAppTime_j(\mathrm{JS},\mathrm{HC})\nonumber\\
SoloRunTime(\mathrm{JS},P_{total}) &=& \sum_{\mathrm{J_j}\in \mathrm{JS}} SoloRunAppTime_j(P_{total})\nonumber\\
CoRunAppTime_j(\mathrm{JS},\mathrm{HC}) &=& Slowdown_j(\mathrm{JS},\mathrm{HC}) \cdot SoloRunAppTime_j(P_{max}) \nonumber\\
SoloRunAppTime_j(P_{total}) &=& Slowdown_j(\{\mathrm{J_j}\},\{R^c_{max}, R^g_{max}, OptP^c_j, OptP^g_j\}) \nonumber\\
& & \cdot SoloRunAppTime_j(P_{max}) \nonumber
\end{eqnarray}
The parameters and functions used to formulate them are summarized in Table~\ref{table:refs2}. 
The first equation denotes that the total execution time when co-scheduling a job set ($\mathrm{JS}$) is determined by the longest execution time in the set.  
The second equation represents that the total execution time of time-shared scheduling is simply the sum of the solo-run execution time of the jobs in the set. 
The third equation shows that the execution time of a co-scheduled job ($CoRunAppTime_j$) is equal to the slowdown ($Slowdown_j$) multiplied by the solo-run execution time without power capping ($SoloRunAppTime_j(P_{max})$). 
In the fourth one, the performance degradation caused by power capping for a solo run can be described by using the same slowdown function used in the third equation. 
In this paper, the solo-run execution time without power capping is given by the associated profile, and we predict the slowdown part in those last two equations. 


\begin{table}[t]
{
\scriptsize
\centering
\begin{tabular}{|M{0.32\linewidth}||M{0.65\linewidth}|}
    \hline
    Parameter or Function & Remarks \\
    \hline\hline
    $CoRunAppTime_j(\mathrm{JS}$, $\mathrm{HC})$ & The execution time of $j$th job in a given job set ($\mathrm{JS}$) when co-scheduling $\mathrm{JS}$ under a given hardware setup ($\mathrm{HC}$) \\
    \hline
    $SoloRunAppTime_j(P_{total})$ & The execution time of $j$th job in a given job set ($\mathrm{JS}$) when its exclusive solo run under a power cap ($P_{total}$) \\
    \hline
    $Slowdown_j(\mathrm{JS},\mathrm{HC})$ & The slowdown ratio of $j$th job in a given job set ($\mathrm{JS}$) caused by co-scheduling $\mathrm{JS}$ under a given hardware setup ($\mathrm{HC}$) \\
    \hline
    $\mathrm{J_j}$ & $j$th job in a given job set ($\mathrm{JS}$) --- $\mathrm{JS} = \{\mathrm{J_1}, \mathrm{J_2}, \cdots\}$ \\
    \hline
    $R^{*}_{max} (* = c/g)$ & The maximum resource allocation on CPU/GPU to a given job\\
    \hline    
    $OptP^{*}_j (* = c/g)$ & The optimal power cap set up on CPU/GPU for $j$th job in a set when exclusive solo run ($OptP^c_j+ OptP^g_j=P_{total}$) \\
    \hline
    $P_{max}$ & The maximum total power cap or TDP ($P_{total} \leq P_{max}$) \\
    \hline
    $F_k^j$ & $k$th parameter to characterize the features of $\mathrm{J_j}$, given by hardware performance counters on both CPU and GPU \\
    \hline
\end{tabular}
}
\caption{Definitions of Parameters or Functions to Formulate $CoRunTime()$/$SoloRunTime()$}
\label{table:refs2}
\end{table}

\begin{figure}[t]
  \centering
  \includegraphics[width=0.75\linewidth]{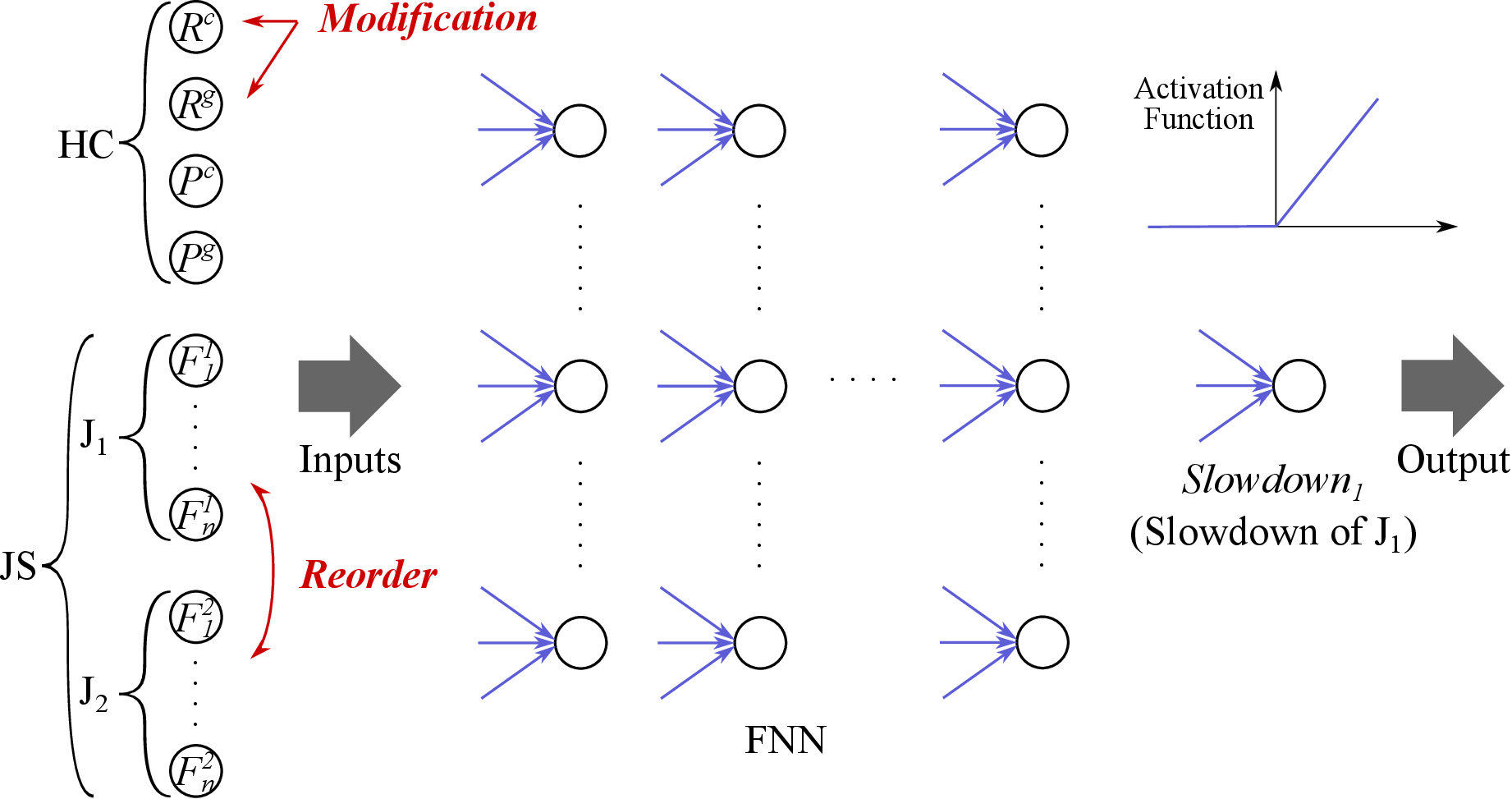}
  \caption{General Structure of Our Performance Modeling ($C=2$)}
  \label{fig:network}
\end{figure}

\vspace{-10pt}
\subsubsection{Performance Modeling:}
Figure~\ref{fig:network} illustrates the general structure to model the slowdown function provided above. 
Here, we utilize a simple feedforward neural network (FNN) to estimate the slowdown for the first job ($\mathrm{J_1}$) in the job set ($\mathrm{JS}$) when co-scheduling. We regard the slowdown as a function of the job features ($F^j_k$) of all the co-located jobs as well as the hardware configuration ($\mathrm{HC}$) to assess various factors such as scalability, interference, and resource allocations. 
The job features here are the hardware performance counters collected from both the CPU and GPU during the profile run described in Section~\ref{solution-overview}. 
The exact definitions for the job features used in our evaluation are listed in Section~\ref{evaluation-setup}. 
As for the slowdowns of the other co-located job(s), we simply reorder or replace the input locations (i.e., exchange the location between $\mathrm{J_1}$ and $\mathrm{J_j}$) and modify the resource allocation parameters ($R_*$) accordingly so that the allocations are associated with the new job order. 
Further, we also utilize the model to estimate the impact of power capping on solo-run performance. To do so, we simply designate $\mathrm{HC}=\{R^c_{max}, R^g_{max}, OptP^c_j, OptP^g_j\}$ as previously mentioned and set all the job features other than the first job to zero in the model inputs. 
The detailed network configurations such as the exact inputs, the layer setups, the activation function, or the loss function are described in Section~\ref{evaluation-setup}.

\subsection{Hardware Setup Optimization for a Given Job Set}
By using the performance model provided above, we optimize the hardware configuration parameters ($\mathrm{HC}$) for a given job set ($\mathrm{JS}$) when co-scheduling.
Here, we attempt to pick up the best hardware configuration ($\mathrm{HC}$)
from all the possible configurations so as to 
minimize $CoRunTime(\mathrm{JS}, \mathrm{HC})$. 
In this study, we simply utilize the exhaustive search, i.e., testing all the possible $\mathrm{HC}$ for the model inputs and choosing one that minimizes $CoRunTime$ for the given job set ($\mathrm{JS}$). 
This is because the number of all the possible setups for $\mathrm{HC}$ on our target platform (or other systems available today) is limited as described later in Section~\ref{evaluation-setup}. 
If the configuration space would explode in future systems, applying heuristic algorithms (e.g., hill climbing) would be a promising option. 
In addition, we select the pair of ($OptP^c_j$, $OptP^g_j$) for each job ($\mathrm{J_j}$) in a given job set so as to obtain $SoloRunTime(\mathrm{JS}, P_{total})$, for which we also explore in an exhaustive manner under the constraint of $OptP^c_j$ + $OptP^g_j = P_{total}$.

\begin{figure}[t]
  \centering
  \includegraphics[width=0.55\linewidth]{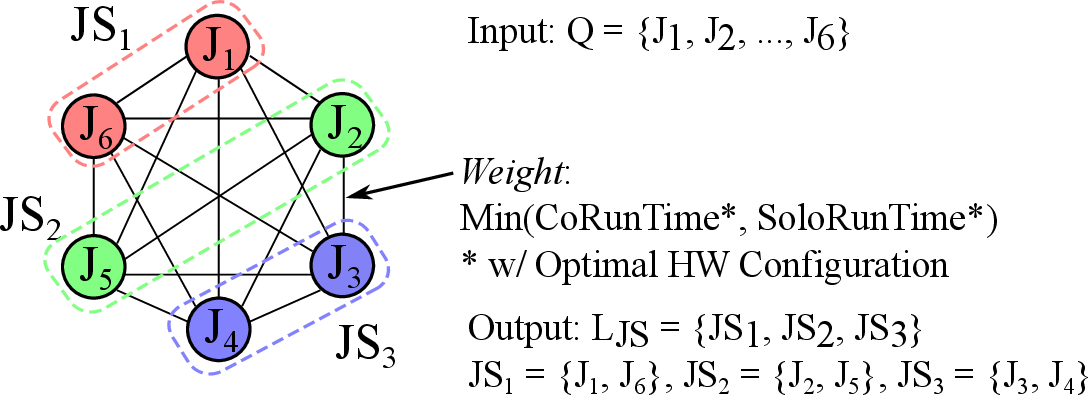}
  \caption{Overview of Graph-based Job Sets Creation ($W=6$, $C=2$)}
  \label{fig:graph}
\end{figure}

\begin{algorithm}[t]
\scriptsize
\nonl\textbf{Inputs:} $\mathrm{Q}=\{J_1, \cdots, J_{W}\}$, $P_{total}$, $\mathrm{SP}=\{C=2, W\}$\\
\nonl\textbf{Outputs:} $\mathrm{L_{JS}} = \{\mathrm{JS_1}, \mathrm{JS_2}, \cdots\}$, $\mathrm{L_{HC}} = \{\mathrm{HC_1}, \mathrm{HC_2}, \cdots\}$\\
\nonl\hrulefill\\
\tcc{Initialization}
$\mathrm{L_{JS}}\gets\varnothing$; $\mathrm{L_{HC}}\gets\varnothing$; 

Vortexes$\gets\mathrm{Q}$; Edges$\gets\varnothing$;  Weights$\gets\varnothing$; HWConfigs$\gets\varnothing$;  CoRunFlags$\gets\varnothing$;

\tcc{Graph creation}
\For{$i=1 \rightarrow W$}
{
    \For{$j=i+1 \rightarrow W$}
    {
        Edges.push\_back(\{$J_i$,$J_j$\}); \tcp{Append this job set}
        
        (HCco, $CoRunTime$) $\gets$ GetOptimalCoRunHWConfig($J_i$,$J_j$);

        (HCsolo1, HCsolo2, $SoloRunTime$) $\gets$ GetOptimalSoloRunHWConfig($J_i$,$J_j$);

        \tcc{Append the weight and the HW config (incl. co-run or solo-runs) that minimizes time for this job set}
    	\uIf{$CoRunTime \leq SoloRunTime$}{
            Weights.push\_back($CoRunTime$); 
            CoRunFlags.push\_back(1); 
            HWConfigs.push\_back(\{HCco\}); 
    	}
	    \Else{
            Weights.push\_back($SoloRunTime$); 
            CoRunFlags.push\_back(0); 
            HWConfigs.push\_back(\{HCsolo1, HCsolo2\}); 
	    }

    }
    
}
\tcc{Job sets decision w/ Edmonds' Algorithm}
$\mathrm{L'_{JS}}$ $\gets$ EdmondsAlgorithm(Vortexes, Edges, Weights); 

$\mathrm{L'_{HC}}$ $\gets$ PickupSets(HWConfigs, $\mathrm{L'_{JS}}$); \tcp{Pick the associated HW setups /w $\mathrm{L'_{JS}}$}
$\mathrm{L_{Flag}}$ $\gets$ PickupSets(CoRunFlags, $\mathrm{L'_{JS}}$); \tcp{Create a co-/solo-run flag list} 

\tcc{Divide sets in $\mathrm{L'_{JS}}$/$\mathrm{L'_{HC}}$ if solo-run execution is better than co-scheduling}
\While{$\mathrm{L_{Flag}} \neq \varnothing$}
{
        $Flag\gets\mathrm{L_{Flag}.pop\_front()}$; 
        $\mathrm{JS}\gets\mathrm{L'_{JS}.pop\_front()}$; 
        $\mathrm{HC}\gets\mathrm{L'_{HC}.pop\_front()}$; 
        
    	\uIf{$Flag=1$}{
            $\mathrm{L_{JS}.push\_back(JS)}$; 
            $\mathrm{L_{HC}.push\_back(HC)}$; 
    	}
	    \Else{
            \While{$\mathrm{JS} \neq \varnothing$}
            {
                $J\gets\mathrm{JS.pop\_front()}$;
                $HCsolo\gets\mathrm{HC.pop\_front()}$; 
                
                $\mathrm{L_{JS}.push\_back(\{J\})}$; 
                $\mathrm{L_{HC}.push\_back(\{HCsolo\})}$;
            }
	    }
}

return ($\mathrm{L_{JS}}$, $\mathrm{L_{HC}}$);

\caption{Job Scheduling Procedure ($C=2$)}\label{job-algorithm}
\end{algorithm}

\subsection{Job Sets Selection}
We then make scheduling decisions using the above hardware setup optimization functionality based on the results of our performance model. 
We regard the job co-scheduling problem as a minimum weight perfect matching problem and solve it using Edmonds' algorithm~\cite{Edmond}. 
Figure~\ref{fig:graph} depicts the overview of the solution. 
In the figure, the vertices represent the jobs in the queue ($\mathrm{Q}$ = $\{\mathrm{J_1}, \cdots, \mathrm{J_W}\}$), and the weights represent the minimum execution time for the associated job sets. 
To obtain each weight, we estimate both of the best $CoRunTime$ and $SoloRunTime$ for each edge (or job pair) by using the model-based hardware configuration optimization described above, and choose one from them so that the execution time is minimized. 
Then, by using the graph, we create the list of job sets ($\mathrm{L_{JS}}$ = $\{\mathrm{JS_1}, \mathrm{JS_2}, \cdots\}$) that includes all jobs in the queue in a mutually exclusive and collectively exhaustive manner, while minimizing the sum of the weights of $\mathrm{L_{JS}}$. 
This is a well-known minimum weight perfect matching problem and is identical to the optimization problem defined in Section~\ref{prob-def} except that a job set can be executed in the time-sharing manner, which we can easily convert to meet the problem definition in Section~\ref{prob-def} by simply dividing such a job set into multiple job sets, all of which include only one job.
The Edmonds' algorithm provides the optimal solution with polynomial time complexity, particularly when the scheduling parameter set ($\mathrm{SP}$) meets both of the following conditions: (1) $W$ is an even number; and (2) $C$ is equal to 2~\cite{Edmond}. 
For the former, we simply set the window size to an even number, and as for the latter, we focus on $C=2$ to limit the complexity as described before. 
Note that a more precise version of the solution is described in Algorithm~\ref{job-algorithm}.

\begin{table}[b]
{
\scriptsize
\vspace{-5pt}

    \caption{Evaluation System Specifications}
    \label{table:specs}
    \vspace{-20pt}
\begin{center}
\begin{tabular}{|M{0.25\linewidth}||M{0.70\linewidth}| } 
    \hline
    Name & Remarks \\
    \hline
    \hline
    CPU & AMD Ryzen Threadripper 2990WX, 32 cores \\
    \hline
    Main Memory & DDR4 2933 MT/s x4ch, 64GB (Total)\\
    \hline
    GPU & NVIDIA A100 40GB PCIe, 8GPCs \\
    \hline
    Operating system & Ubuntu 20.04.4 LTS, Kernel Version 5.4.0-120-generic \\
    \hline
    Compiler and drivers & GCC/G++ Version: 9.4.0, CUDA Version: 11.6, Driver Version: 510.73.08 \\
    \hline
\end{tabular}
\end{center}
}
\vspace{-5pt}
{
\scriptsize
\caption{Power Cap and Partitioning Setups}\label{search-space}
\vspace{-20pt}
\begin{center}
\begin{tabular}{ |M{0.25\linewidth}||M{0.70\linewidth}| } 
\hline
 Variable & Selections \\\hline\hline
 $P_{total}$ / $P_{max}$ &  350, 400 [W] / 500 [W] \\\hline
 $P_{*}^{c}$ &  100, 125, 150, 175, 200, 225, 250(max) [W] \\\hline
 $P_{*}^{g}$ &  150, 175, 200, 225, 250(max) [W] \\\hline
 $R_{*}^{c}$ &  (\# of cores for $\mathrm{J_1}$, \# of cores for $\mathrm{J_2}$): (2,30), (8,24), (16,16), (24,8), (30,2) (= co-runs), (32,0) (= solo-run, $R_{max}^{c}$) \\\hline
 $R_{*}^{g}$ &  (\# of GPCs for $\mathrm{J_1}$, \# of GPCs for $\mathrm{J_2}$): (3,4), (4,3) (= co-runs), (8,0) (= solo-run, $R_{max}^{g}$) \\\hline
\end{tabular}
\end{center}
}
\end{table}

\section{Evaluation}
\subsection{Evaluation Setup}\label{evaluation-setup}

\subsubsection{Environment}
For our evaluation, we use the platform summarized in Table~\ref{table:specs}. 
Our approach is applicable when both the CPU and GPU are capable of both resource partitioning and power capping. 
This is usually the case for most of the commercial CPUs today, and we utilize an NVIDIA A100 GPU card that supports the MIG feature and power capping~\cite{mig2}. 


Table~\ref{search-space} summarizes the resource partitioning and power capping settings we explore in this evaluation. 
We allocate CPU cores in a compact fashion, i.e., physically adjacent cores are assigned to the same program. 
We partition the GPU into 3GPCs/4GPCs or 4GPCs/3GPCs, on which low level memory hierarchies including L2 caches and memory modules are shared across all the GPCs\footnote{One GPC must be disabled when using MIG. Other partitioning options such as 1GPC/6GPCs or 2GPCs/5GPCs are not supported. We first create one GI with 7GPCs and then create CIs consisting of 3GPCs/4GPCs inside of it~\cite{mig-icppw,mig2}. }. 
To collect performance counter values when profiling, we utilize Linux perf~\cite{perf} command for the CPU and NSight Compute~\cite{nsight-compute} for the GPU. By using these profiling frameworks, we collect the performance counter values listed in Table~\ref{counters}. 
The definitions of these performance counters are the same as those shown in the tools.

\begin{table}[b]
{
\scriptsize
\caption{Collected Performance Counters ($\mathbf{F}$)}\label{counters}
\vspace{-20pt}
\begin{center}
\begin{tabular}{ |M{0.15\linewidth}||M{0.8\linewidth}| } 
\hline
Component & Counters and Definitions
\\\hline\hline
CPU & $F_1^*$ = cpu-util, $F_2^*$ = context-switches, $F_3^*$ = page-faults, $F_4^*$ = IPC, $F_5^*$ = stalled-cycles, $F_6^*$ = branch-misses, $F_7^*$ = L1-dcache-load-misses, $F_8^*$ = L1-icache-load-misses, $F_9^*$ = dTLB-load-misses, $F_{10}^*$ = iTLB-load-misses
\\\hline 
GPU & $F_{11}^*$ = Memory\lbrack\%\rbrack, $F_{12}^*$ = DRAM Throughput\lbrack\%\rbrack, $F_{13}^*$ = TEX cache Throughout\lbrack\%\rbrack, $F_{14}^*$ = LLC Throughput\lbrack\%\rbrack, $F_{15}^*$ = Compute\lbrack\%\rbrack,  $F_{16}^*$ = Waves per SM, $F_{17}^*$ = Achieved Occupancy\lbrack\%\rbrack, and $F_{18}^*$ = Warps per SM
\\\hline 
\end{tabular}
\end{center}
}
{
\scriptsize
\caption{Tested Job Mixes}\label{job-mix}
\vspace{-20pt}
\begin{center}
\begin{tabular}{ |M{0.15\linewidth}||M{0.8\linewidth}| } 
\hline
Name & Job Mix
\\\hline\hline
JobMix1 & $\mathrm{Q}=$\{gaussian, lud, pathfinder, streamcluster\}, $C=2$, $W=4$ 
\\\hline 
JobMix2 & $\mathrm{Q}=$\{gaussian, srad, hotspot, pathfinder, lavaMD, matvec\}, $C=2$, $W=6$
\\\hline 
JobMix3 & $\mathrm{Q}=$\{gaussian, srad, hotspot, lud, pathfinder, lavaMD, streamcluster, matvec\}, $C=2$, $W=8$
\\\hline 
\end{tabular}
\end{center}
}
\end{table}

\vspace{-10pt}
\subsubsection{Benchmarks and Dataset}
We use the Rodinia benchmarks~\cite{rod1}, which is a well-known benchmark suite widely-used for various heterogeneous computing studies, as well as a synthetic compute-intensive dense matrix-vector multiplication program (\texttt{matvec}). In particular, from the Rodinia benchmark suite, we pick up seven programs that utilize both CPU and GPU extensively/cooperatively. 
Further, the \texttt{matvec} program uses both CPU and GPU in a cooperative manner, i.e., a part of the computation is offloaded to GPU and the rest is performed on CPU. 
We then create three different job queues (\textit{JobMix1}, \textit{JobMix2}, and \textit{JobMix3}) with different window sizes ($W$) ranging from 4 to 8. 
The programs in the queues are selected mutually-exclusively (and randomly for \textit{JobMix1}/\textit{JobMix2}) from the eight benchmarks. 

We then generate the training/validation/test datasets by using the benchmarks. 
More specifically, we randomly select 8x2=16 job pairs out of all the possible ${}_8 C_2=28$ pairs and measure the co-scheduling slowdowns for each of them while testing 100 different hardware setups that is identical to all the co-run hardware setups that meet $P_{total}$ = 350 or 400 [W] in Table~\ref{search-space}. 
To validate the performance model, we divide the dataset in the following way: the first 12 pairs multiplied by 100 hardware configurations (= 2,400 data points) are used for the training and validation; and the rest of the 800 data are used for the inference testing. 
Note the above division process is based on random pair selections. 
The training and validation here are corresponding to the offline procedure shown in Figure~\ref{fig:workflow} in Section~\ref{solution-overview}. 

\vspace{-10pt}
\subsubsection{Neural Network Architecture and Training}
Table~\ref{nn-params} lists our neural network architecture and training setups based on the general structure described in Section~\ref{modeling}. 
In our neural network, all the inputs are normalized between 0 and 1 (including the hardware configuration) in order to equalize the significance of them, which ultimately helps the convergence. 
To normalize the resource partitioning states ($R^*_i$), 
we simply pick the first element that represents the number of core or GPC allocation to the first job ($\mathrm{J_1}$), and then divide it by the maximum number of the resource allocation ($32$ for cores and $8$ for GPCs in our environment). 
We set up two hidden layers to well recognize the patterns in the input values, which is better than relying on one single hidden layer for this purpose. 
The rectified linear activation function is applied to all the layers except for the input layer, and all the neurons in both the hidden layers and the output have biases. 
The input layer is fully connected with the first hidden layer in order to use the model while re-ordering the job inputs (see also Section~\ref{modeling}). 
In our Python implementation, the training with the dataset described above takes only few minutes, and the slowdown estimations for all the jobs in a job set takes only 1.17 ms in total.

\begin{table}[b]
{
\scriptsize
\caption{Model and Training Setups}\label{nn-params}
\vspace{-20pt}
\begin{center}
\begin{tabular}{ |M{0.15\linewidth}||M{0.8\linewidth}| } 
\hline
Type & Parameter List
\\\hline\hline
Model & \lbrack\textbf{Input layer}\rbrack = 4 HW config states ($\mathrm{HC}$) + 18 HW counters ($\mathrm{J_1}$) + 18 HW counters ($\mathrm{J_2}$); 
\lbrack\textbf{\# of hidden layers}\rbrack = 2; \lbrack\textbf{\# of neurons in each hidden layers}\rbrack = 18 (= \# of HW counters); \lbrack\textbf{Layer connection}\rbrack = Fully connected; \lbrack\textbf{Activation function}\rbrack = Rectified Linear
\\\hline 
Training & \lbrack\textbf{Learning rate}\rbrack = 0.001; \lbrack \textbf{Batch size}\rbrack = 4; \lbrack\textbf{Optimizer}\rbrack = Stochastic Gradient Descent; \lbrack\textbf{\# of epochs}\rbrack = 200; \lbrack\textbf{Loss function}\rbrack = Mean Square Error
\\\hline 
\end{tabular}
\end{center}
}
\end{table}

\subsection{Experimental Results}

\begin{figure}[t]
\begin{center}
\begin{tabular}{c}
\begin{minipage}{0.5\hsize}
  \begin{center}
    \includegraphics[width=\linewidth]{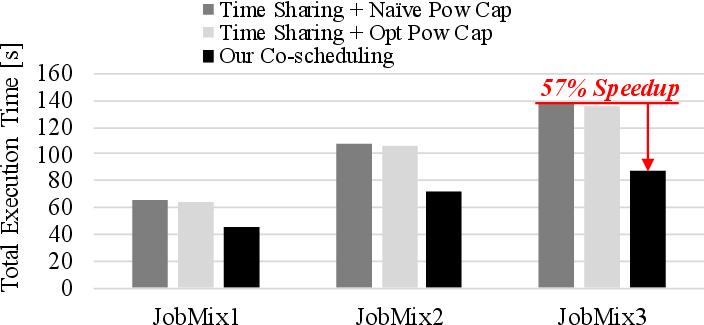}
    \caption{Execution Time Comparison ($P_{total}=350[W]$)}
    \label{perf-comp-350w}
  \end{center}
\end{minipage}

\begin{minipage}{0.5\hsize}
  \begin{center}
    \includegraphics[width=\linewidth]{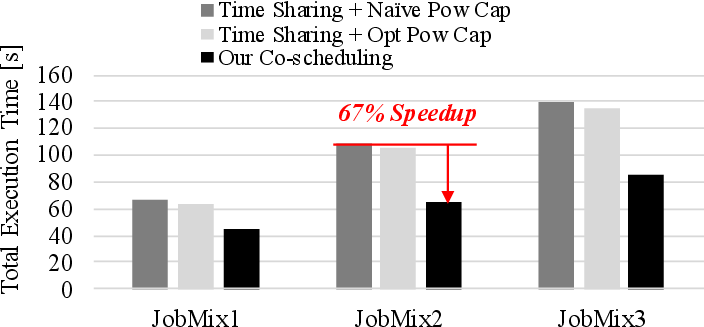}
    \caption{Execution Time Comparison ($P_{total}=400[W]$)}
    \label{perf-comp-400w}
  \end{center}
\end{minipage}

\end{tabular}
\end{center}
\end{figure}

\begin{figure}[t]
\begin{center}
\begin{tabular}{c}
\begin{minipage}{0.5\hsize}
  \begin{center}
    \includegraphics[width=\linewidth]{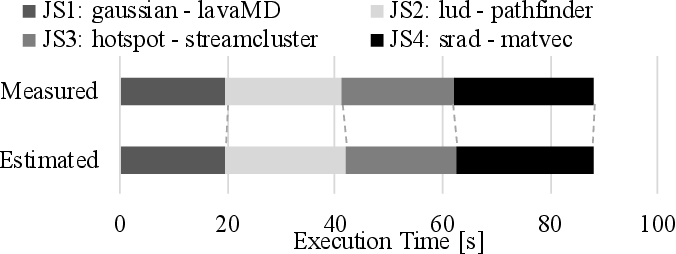}
    \caption{Comparison of Measured vs Estimated Time (\textit{JobMix3}, $P_{total}=350[W]$)}
    \label{sched-accuracy-350w}
  \end{center}
\end{minipage}

\begin{minipage}{0.5\hsize}
  \begin{center}
    \includegraphics[width=\linewidth]{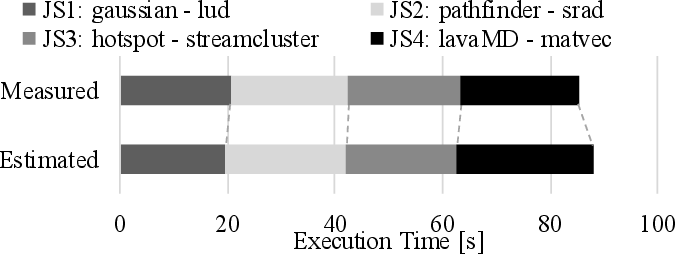}
    \caption{Comparison of Measured vs Estimated Time (\textit{JobMix3}, $P_{total}=400[W]$)}
    \label{sched-accuracy-400w}
  \end{center}
\end{minipage}

\end{tabular}
\end{center}
\end{figure}

Figure~\ref{perf-comp-350w}/\ref{perf-comp-400w} demonstrate the total execution time comparisons across multiple different scheduling and resource management polices for different total power cap setup ($P_{total}=350,400$[W]). 
The vertical axis indicates the total execution time, while the horizontal axis lists job queues (\textit{JobMix1}-\textit{JobMix3}) in both the figures. 
The details of the compared policies listed in the legends are as follows: 
\textit{Time Sharing + Naive Pow Cap} schedules jobs in the time-sharing manner while setting up the power caps to the CPU and GPU equally; \textit{Time Sharing + Opt Pow Cap} also utilizes the time-shared scheduling but the power caps are set to the optimal; and \textit{Our Co-scheduling} schedules jobs and configures the hardware using our proposed approach. 
As shown in these figures, we achieve significant speedups by up to 67.4\% (= (108.8/65.0 - 1)*100) by using our approach compared with \textit{Time Sharing + Naive Pow Cap}. Note the hardware partitioning is done only at the job launches, thus the overhead is negligible here. 
Table~\ref{cosh-res} presents the list of job sets created by our approach for each queue under different power capping. The job set selections can change depending on the total power cap setup, which implies our approach can flexibly deal with hardware environment changes, e.g., with changes in the power supply level. 

\begin{table}[t]
{
\scriptsize
\caption{Lists of Job Sets Created by Our Approach}\label{cosh-res}
\vspace{-20pt}
\begin{center}
\begin{tabular}{ |M{0.10\linewidth}||M{0.85\linewidth}| } 
\hline
$P_{total}$ & Lists of Job Sets ($\mathrm{L_{JS}}$) 
\\\hline\hline
350W & (\textit{JobMix1}): \{gaussian-lud, pathfinder-streamcluster\}, (\textit{JobMix2}): \{gaussian-hotspot, pathfinder-lavaMD, matvec-srad\}, (\textit{JobMix3}): \{lavaMD-gaussian, lud-pathfinder, hotspot-streamcluster, matvec-srad\} 
\\\hline 
400W & (\textit{JobMix1}): \{gaussain-lud, pathfinder-streamcluster\}, (\textit{JobMix2}): \{hotspot-lavaMD, srad-pathfinder, matvec-gaussian\}, (\textit{JobMix3}): \{gaussian-lud, srad-pathfinder, hotspot-streamcluster, lavaMD-matvec\} 
\\\hline 
\end{tabular}
\end{center}
}
\end{table}


We then compare the measured and estimated execution time (excluding online scheduling time) for different power cap setups in Figure~\ref{sched-accuracy-350w}/\ref{sched-accuracy-400w}. 
The X-axis indicates the accumulated execution time of all the co-scheduled job sets created from \textit{JobMix3} by using our approach. 
As shown in the figure, the estimated execution times are close to the measured ones, and the total estimation error is only 0.4\% or 3.1\% for $P_{total}=350$ or 400, respectively. 
Note that our approach achieves closer performance to the optimal as the error becomes smaller. 
This is because the Edmonds' algorithm returns the optimal scheduling job sets if the performance estimation is 100\% accurate. 


\begin{figure}[t]
\begin{center}
\begin{tabular}{c}
\begin{minipage}{0.33\hsize}
  \begin{center}
    \includegraphics[width=\linewidth]{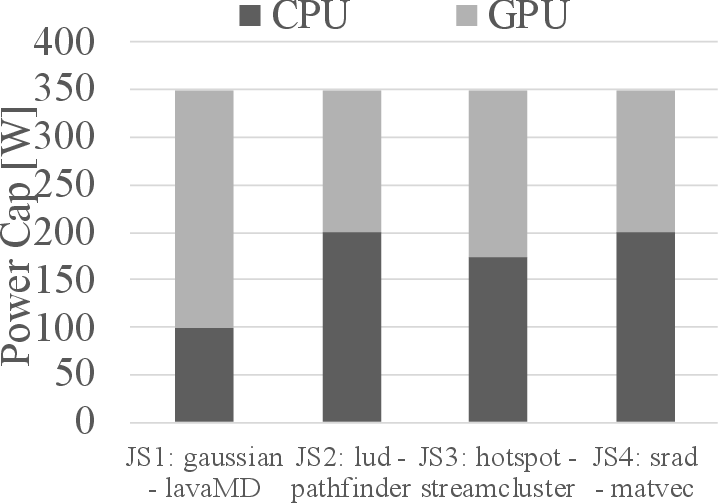}
    \caption{Power Cap Setups (JobMix3, $P_{total}=350[W]$)}
    \label{pow-cap-350w}
  \end{center}
\end{minipage}

\begin{minipage}{0.33\hsize}
  \begin{center}
    \includegraphics[width=\linewidth]{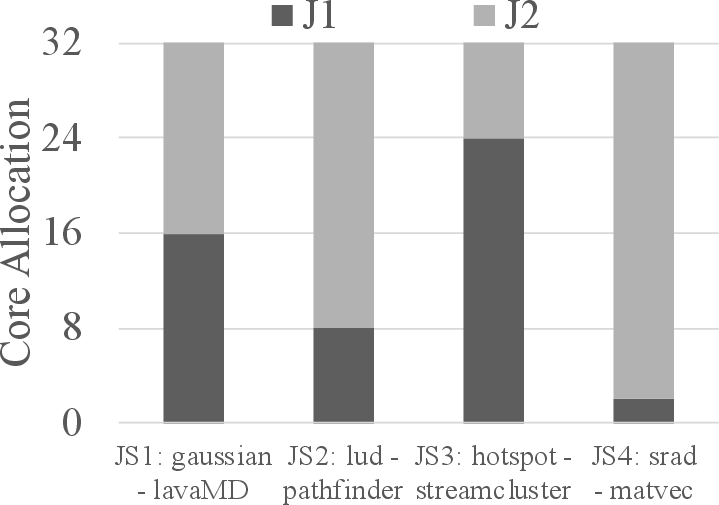}
    \caption{Core Allocation (JobMix3, $P_{total}=350[W]$)}
    \label{core-alloc-350w}
  \end{center}
\end{minipage}

\begin{minipage}{0.33\hsize}
  \begin{center}
    \includegraphics[width=\linewidth]{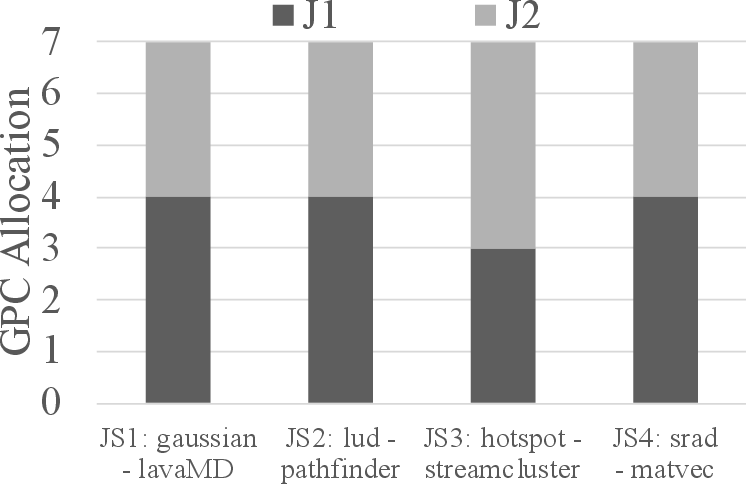}
    \caption{GPC Allocation (JobMix3, $P_{total}=350[W]$)}
    \label{gpc-alloc-350w}
  \end{center}
\end{minipage}

\end{tabular}
\end{center}
  \vspace{-5pt}
\end{figure}

Finally, we demonstrate the hardware setup decisions made by our scheduler in Figure~\ref{pow-cap-350w}/\ref{core-alloc-350w}/\ref{gpc-alloc-350w}, in particular, for \textit{JobMix3} under the total power cap of $P_{total}=350$[W]. 
The X-axis indicates the job sets created from \textit{JobMix3} by our approach, while the Y-axis represents the breakdown of power caps, core allocations, or GPC allocations in Figure~\ref{pow-cap-350w},~\ref{core-alloc-350w},~or~\ref{gpc-alloc-350w}, respectively. 
As shown in these figures, these hardware knobs are set very differently in accordance with the characteristics of co-located jobs, including the task size on CPU/GPU, the compute/memory intensity, and the interference on shared resources. 
As our performance modeling can recognize these features well based on the corresponding hardware performance counters and the well-structured neural network, our approach achieves the significant performance improvement by up to 67\%. 
\vspace{-10pt}
\section{Conclusion}
\label{Conclusion}
\vspace{-10pt}

In this paper, we targeted co-scheduling, resource partitioning, and power capping comprehensively for CPU-GPU heterogeneous systems and proposed an approach to optimize them, which consists of performance modeling and a graph-based scheduling algorithm. 
We demonstrated how a machine learning model, namely a neural network, can successfully be used to predict the performance of co-scheduled applications, while using the application characteristics and partitioning/power states as inputs. 
We then moved on to the application pair selections where we successfully applied Edmond's algorithm to determine the mathematically optimal pairing. 
The experimental result using a real system shows that our approach improves the system throughput by up to 67\% compared with a time-sharing-based scheduling with a naive power capping that evenly distributes power budgets on CPU/GPU.

\vspace{-5pt}
\subsubsection{Acknowledgements} 
This work has received funding under the European Commission's EuroHPC and H2020 programmes under grant agreement no. 956560 and was supported by the NVIDIA Academic Hardware Grant Program. 

%
%
{
\bibliographystyle{splncs04}
\renewcommand{\baselinestretch}{0.98}
\bibliography{ref.bib}
}

\end{document}